\documentclass[english,aps,prl,reprint,superscriptaddress,showpacs]{revtex4-1}
\usepackage[T1]{fontenc}
\usepackage[latin9]{inputenc}
\usepackage{amsmath}
\usepackage{amssymb}
\usepackage{graphicx}
\usepackage{esint}

\makeatletter
\@ifundefined{textcolor}{}
{%
 \definecolor{BLACK}{gray}{0}
 \definecolor{WHITE}{gray}{1}
 \definecolor{RED}{rgb}{1,0,0}
 \definecolor{GREEN}{rgb}{0,1,0}
 \definecolor{BLUE}{rgb}{0,0,1}
 \definecolor{CYAN}{cmyk}{1,0,0,0}
 \definecolor{MAGENTA}{cmyk}{0,1,0,0}
 \definecolor{YELLOW}{cmyk}{0,0,1,0}
 }

\makeatother

\usepackage{babel}
\begin{document}

\title{Strong and weak chaos in nonlinear networks with time-delayed couplings}

\author{Sven Heiligenthal}

\email{sven.heiligenthal@physik.uni-wuerzburg.de}

\affiliation{Institute of Theoretical Physics, University of Würzburg, Am Hubland,
97074 Würzburg, Germany}

\author{Thomas Dahms}

\affiliation{Institute of Theoretical Physics, Technical University of Berlin, Hardenbergstraße
36, 10623 Berlin, Germany}

\author{Serhiy Yanchuk}

\affiliation{Institute of Mathematics, Humboldt University of Berlin, Unter den
Linden 6, 10099 Berlin, Germany}

\author{Thomas Jüngling}

\affiliation{Institute of Theoretical Physics, University of Würzburg, Am Hubland,
97074 Würzburg, Germany}

\author{Valentin Flunkert}

\affiliation{Institute of Theoretical Physics, Technical University of Berlin, Hardenbergstraße
36, 10623 Berlin, Germany}

\author{Ido Kanter}

\affiliation{Department of Physics, Bar-Ilan University, Ramat-Gan 52900, Israel}

\author{Eckehard Schöll}

\affiliation{Institute of Theoretical Physics, Technical University of Berlin, Hardenbergstraße
36, 10623 Berlin, Germany}

\author{Wolfgang Kinzel}

\affiliation{Institute of Theoretical Physics, University of Würzburg, Am Hubland,
97074 Würzburg, Germany}

\date{\today}
\begin{abstract}
We study chaotic synchronization in networks with time-delayed coupling.
We introduce the notion of strong and weak chaos, distinguished by
the scaling properties of the maximum Lyapunov exponent within the
synchronization manifold for large delay times, and relate this to
the condition for stable or unstable chaotic synchronization, respectively. 
In simulations
of laser models and experiments with electronic circuits, we identify
transitions from weak to strong and back to weak chaos upon monotonically
increasing the coupling strength.
\end{abstract}


\maketitle
The cooperative behavior of a system of interacting units is of fundamental
interest in nonlinear dynamics. Such complex networks have a wide
range of interdisciplinary applications ranging from neural networks
to coupled lasers \cite{Albert2002,*Boccaletti2006,*Arenas2008,*Song2010}.
Typically, these units interact by transmitting information about
their state to their partners, and in many applications the transmission
time is larger than the time scales of the individual units. Thus,
networks with time-delayed couplings are a focus of active research
\cite{Lakshmanan2010,*Just2010}.

Time-delayed feedback can produce dynamical instabilities which may
lead to deterministic chaos \cite{Farmer1982,Lepri1994,Giacomelli1996}.
Even a scalar differential equation with time-delayed feedback has
an infinite-dimensional phase space which favors chaotic solutions.
In physics, a single semiconductor laser produces a chaotic signal
when its laser beam is reflected back into its cavity by an external
mirror. Networks of nonlinear units may, similarly, become chaotic due to 
time-delayed coupling of the nodes.
For networks of identical units, one often observes chaos synchronization.
Even if the delay times are very long, the units may synchronize onto
a common chaotic trajectory without time shift (zero-lag synchronization)
\cite{Klein2006,*Fischer2006,Flunkert2010}. Other kinds of synchronization
are possible, as well, like phase, achronal, anticipated and generalized
synchronization, but here we only consider zero-lag synchronization.
Chaos synchronization is being discussed in the context of secure
communication \cite{Argyris2005,*Kanter2008}.

In this letter we investigate networks with time-delayed couplings
in the limit of large delay times \cite{Flunkert2010,Englert2011},
and show that transitions between two kinds of chaos, namely \emph{strong}
and \emph{weak} chaos, can be induced by changing the coupling strength.
For strong chaos the largest Lyapunov exponent (LE) is of the order
of the inverse time scales of the individual units and independent
of the delay time, hence two nearby trajectories separate very quickly.
For weak chaos, however, the LE is of the order of the inverse delay
time, hence nearby trajectories separate very slowly. We show that
these two types of chaos possess very different synchronization properties:
Networks with strong chaos cannot synchronize, whereas for weak chaos,
networks can synchronize if the product of the LE and the delay time
is sufficiently small compared to the eigenvalue gap of the coupling
matrix, i.\,e. the difference between the row sum and its largest
transversal eigenvalue.

We illustrate our general findings by the example of a semiconductor
laser network modeled by the Lang-Kobayashi (LK) rate equations, and
by experiments on chaotic electronic circuits.

We consider networks of $N$ identical units with variables $\mathbf{x}_{i}(t)\in\mathbb{R}^{n}$,
$i=1,\ldots,N$, which obey the equations
\begin{equation}
\dot{\mathbf{x}}_{i}(t)=\mathbf{F}\!\left[\mathbf{x}_{i}(t)\right]+\sigma\sum_{j}G_{ij}\,\mathbf{H}\!\left[\mathbf{x}_{j}(t-\tau)\right]\;.\label{E1}
\end{equation}
The nonlinear function $\mathbf{F}$ describes the local dynamics
of the individual units. The units are connected by the coupling matrix
$G=\{G_{ij}\in\mathbb{R}\}$, which describes the coupling topology
and the weight of each link. The coupling itself is characterized
by the coupling function $\mathbf{H}$, the delay time $\tau$ and
the strength $\sigma$. We consider coupling matrices $G$ with normalized
row sum ($\sum_{j}G_{ij}=1$), such that complete synchronization
$\mathbf{x}_{i}(t)=\mathbf{s}(t)$ is a solution of Eq.~\eqref{E1},
\begin{equation}
\dot{\mathbf{s}}(t)=\mathbf{F}\!\left[\mathbf{s}(t)\right]+\sigma\,\mathbf{H}\!\left[\mathbf{s}(t-\tau)\right]\;.\label{E2}
\end{equation}
The dynamics $\mathbf{s}(t)$ within the synchronization manifold
(SM) is identical to the dynamics of a single unit with time-delayed
feedback. The LEs of a single unit are obtained from linearizing Eq.~\eqref{E2}
which gives
\begin{equation}
\dot{\boldsymbol{\xi}}(t)=\mathrm{D}F\!\left[\mathbf{s}(t)\right]\mathbf{\boldsymbol{\xi}}(t)+\sigma\,\mathrm{D}H\!\left[\mathbf{s}(t-\tau)\right]\boldsymbol{\xi}(t-\tau)\;.\label{E3}
\end{equation}
The maximum LE $\lambda_{\mathrm{m}}$ of Eq.~\eqref{E3} is a measure
of the chaoticity within the SM. It turns out that it is useful to
consider the maximum LE $\lambda_{0}$ from an integration of the
reduced linear system \cite{Lepri1994}
\begin{equation}
\dot{\boldsymbol{\xi}}(t)=\mathrm{D}F\!\left[\mathbf{s}(t)\right]\boldsymbol{\xi}(t)\;.\label{E4}
\end{equation}
We call this LE $\lambda_{0}$ the \emph{instantaneous Lyapunov exponent}
of the system, since there is no delayed term in the corresponding
variational equation. Note, however, that this should not be confused
with a \emph{finite-time} LE. Furthermore, $\lambda_{0}$ still depends
on the coupling strength $\sigma$, since Eq.~\eqref{E4} contains
the trajectory $\mathbf{s}(t)$. The following results hold in the
limit of large delay times $\tau$: \emph{Weak chaos} occurs if $\lambda_{0}<0$;
in this case $\lambda_{\mathrm{m}}=\eta/\tau$ in the leading order,
where $\eta$ is independent of $\tau$. \emph{Strong chaos} is encountered
if $\lambda_{0}>0$; here $\lambda_{\mathrm{m}}\approx\lambda_{0}$
up to a correction which is exponentially small with respect to $\tau$.
At first, we give a sketch of the proof.

\textbf{Weak chaos} ($\lambda_{0}<0$) --- Let us denote by $X(t,s)$
the fundamental matrix solution \cite{Hale1980} of the instantaneous
linear system Eq.~\eqref{E4}. In the case of weak chaos, it satisfies
$\|X(t,s)\|\le M\,\mathrm{e}^{\lambda_{0}(t-s)}$ with negative $\lambda_{0}$.
Let us split the solution of Eq.~\eqref{E3} into pieces of the length
$\tau$ as follows $\boldsymbol{\xi}_{j}(\theta):=\boldsymbol{\xi}(\theta+\tau\, j)$
with $0\le\theta\le\tau$. Then, $\boldsymbol{\xi}_{j}$ can be expressed
using the variation of constants formula \cite{Hale1980} as follows
\begin{align}
\boldsymbol{\xi}_{j}(\theta) & =X_{j}(\theta,0)\,\boldsymbol{\xi}_{j-1}(\tau)\nonumber \\
 & \quad+\sigma\int_{0}^{\theta}X_{j}(\theta,t')\,\mathrm{D}H[\mathbf{s}(t'-\tau)]\,\boldsymbol{\xi}_{j-1}(t')\,\mathrm{d}t'\label{E5}
\end{align}
where $X_{j}(\theta,t')=X(\theta+\tau\, j,t'+\tau\, j)$. Using the
exponential decrease of $X_{j}(\theta,t')$, it is straightforward
to obtain from Eq.~\eqref{E5} the estimate $\max_{\theta}\|\boldsymbol{\xi}_{j}(\theta)\|\le L\,\max_{\theta}\|\boldsymbol{\xi}_{j-1}(\theta)\|\le L^{j}\,\max_{\theta}\|\boldsymbol{\xi}_{0}(\theta)\|$,
where $L$ is some constant. This immediately implies that the exponential
growth of the solutions is possible only with respect to the slow
time $j=t/\tau$ \cite{Lepri1994,Giacomelli1996}, and, hence, the
maximum LE is scaled by $1/\tau$ in the case when $\lambda_{0}<0$.
Strictly speaking, the constant $L$ depends on $\tau$, since the
properties of the chaotic attractor change with $\tau$. However,
we argue that for large $\tau$ this dependence disappears: A chaotic
attractor is characterized by its {}``skeleton'' of periodic orbits
and in particular by the orbits of low period. The set of periodic
orbits, which exists for low values of $\tau$, reappears generically
also for larger delays \cite{Yanchuk2009}. Thus, we conjecture that
in the limit of large $\tau$ generically all characteristics of the
attractor converge to a limit and in particular $L$ becomes independent
of $\tau$. All experimental and theoretical results about chaos in
delayed systems, for instance in lasers with delayed self-feedback
and optoelectronic oscillators \cite{Callan2009}, support this argument
\cite{Vicente2005}.

\textbf{Strong chaos} ($\lambda_{0}>0$) --- Let us make the coordinate
transformation to the frame diverging exponentially with rate $\lambda_{0},$
i.\,e. $\boldsymbol{\xi}(t)=\mathrm{e}^{\lambda_{0}t}\,\mathbf{u}(t)$.
In the new coordinates, the variational Eq.~\eqref{E3} has the form
\begin{align}
\dot{\mathbf{u}}(t)= & \left(\mathrm{D}F[\mathbf{s}(t)]-\lambda_{0}\, I\right)\mathbf{u}(t)\nonumber \\
 & +\sigma\,\mathrm{e}^{-\lambda_{0}\,\tau}\,\mathrm{D}H\left[\mathbf{s}(t-\tau)\right]\,\mathbf{u}(t-\tau)\;,\label{E6}
\end{align}
where the largest LE of the instantaneous vector field is zero. Applying
the same arguments to the rescaled system \eqref{E6} as in the case
of weak chaos, we conclude that the maximum LE for $\mathbf{u}(t)$
is at most of the order $1/\tau$ and converges to zero for large
delays. Hence, $\lambda_{0}$ approximates $\lambda_{\mathrm{m}}$
for large $\tau$. Numerical calculations (see below) show that, in
fact, the largest LE converges to $\lambda_{0}$ with an error $\mathrm{e}^{-\mu\,\tau}$,
$\mu>0$. The convergence rate, however, is much slower than in the
case of steady states %
\footnote{In the case of steady states, the convergence rate can be estimated
as $|\lambda_{\mathrm{m}}-\lambda_{0}|<\mathrm{const.}\times\mathrm{e}^{-\lambda_{0}\,\tau}$. %
}. Note that the existence of LEs that are independent of $\tau$ has
also been reported in \cite{Lepri1994} for time-discrete maps with
delay. Such exponents have been called anomalous there. They can be
also computed by regarding Eq.~\eqref{E3} as a nonautonomous differential
equation with the delay term acting as {}``stochastic contribution''.
Note further that in most chaotic delayed systems that have been studied,
such as Ikeda and Mackey-Glass oscillators, the local dynamics is a
constant damping, such that these systems only exhibit weak chaos.
We thus propose to investigate delayed systems with strong chaos,
such as lasers with delayed feedback in certain parameter regimes,
further, since these systems may have important applications for instance
as random number generators \cite{Kanter2009}.

\textbf{Consequences} --- We now discuss the consequences of these
results for systems with large delay. For strong chaos, the maximum
LE of the system is already given by the instantaneous term, Eq.~\eqref{E4}.
The coupling strength $\sigma$ contributes only indirectly through
the orbit $\mathbf{s}(t)$. For weak chaos, however, chaos is generated
by the delayed term with strength $\sigma$ in the variational equation,
and the maximum LE is of order $1/\tau$. It is important to note
here that both types of chaos are delay-induced. In fact, the laser
system that we consider exhibits stable continuous wave output without
delayed feedback.

Our theoretical predictions are compared to numerical simulations
of the LK equations modeling a semiconductor laser with optical feedback.
Now the variables $\mathbf{x}_{i}(t)\in\mathbb{R}^{3}$ contain the
real and imaginary parts of the electromagnetic field and the charge
carrier inversion. Details of the equations and parameters can be
found in \cite{Englert2011}. 
\begin{figure}
\includegraphics[height=0.33\columnwidth]{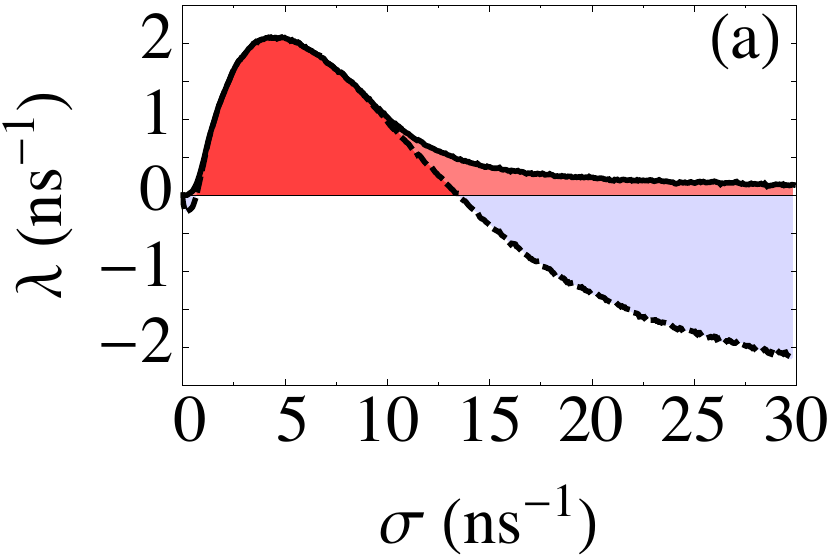}\hspace*{\fill}\includegraphics[height=0.33\columnwidth]{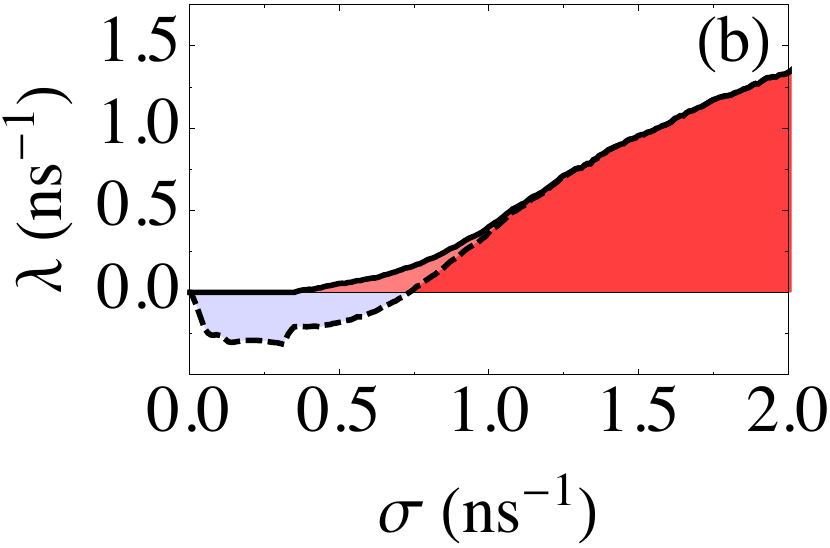}
\caption{\label{F1}(Color online) (a)~Maximum Lyapunov exponents $\lambda_{\mathrm{m}}$
(solid line) and $\lambda_{0}$ (dashed line) of the synchronization
manifold (SM) for $\tau=10\,\mathrm{ns}$ vs. coupling strength $\sigma$.
(b)~Enlarged view for small coupling strengths $\sigma$.}
\end{figure}
Fig.~\ref{F1} shows the two maximum LEs $\lambda_{\mathrm{m}}$
and $\lambda_{0}$ as a function of the coupling strength $\sigma$.
Without coupling, $\sigma=0$, the laser relaxes to a constant intensity,
both LEs are zero and correspond to the Goldstone mode. For a small
coupling, the laser becomes chaotic but the instantaneous LE is negative,
i.\,e., the chaos is weak. With increasing coupling strength, $\lambda_{0}$
increases to positive values. Hence, the laser is strongly chaotic
in some interval of $\sigma$. For higher values of $\sigma$, the
laser is weakly chaotic again. 
\begin{figure}
\includegraphics[height=0.33\columnwidth]{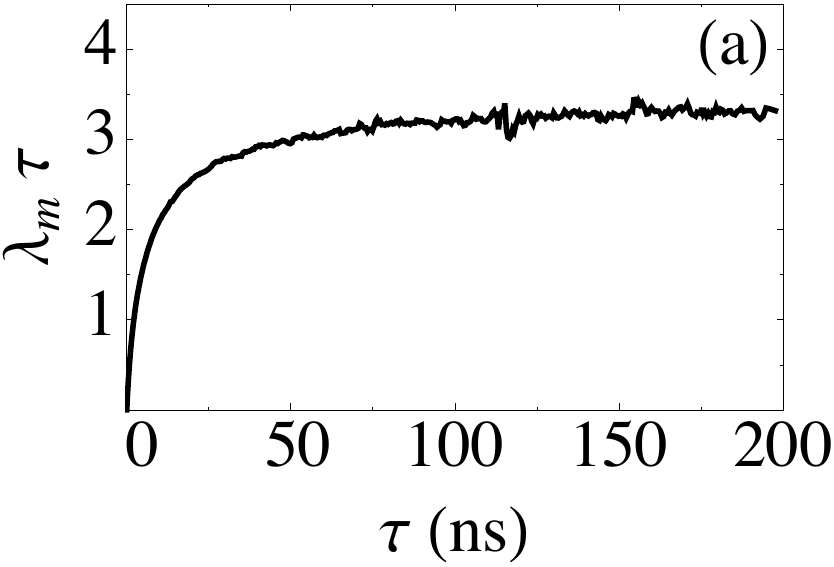}\hspace*{\fill}\includegraphics[height=0.33\columnwidth]{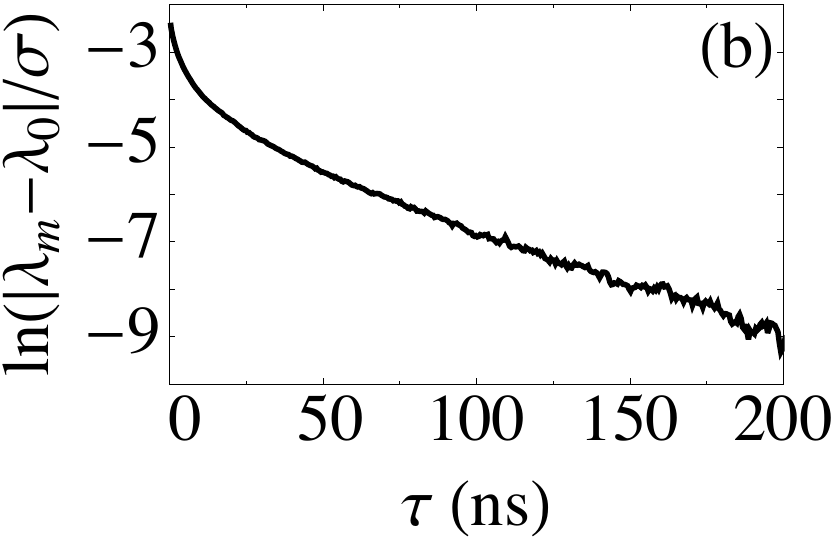}
\caption{\label{F2}(a)~$\lambda_{\mathrm{m}}\,\tau$ for the SM vs. delay
time $\tau$ in the regime of weak chaos ($\sigma=21\,\mathrm{ns}^{-1}$).
(b)~$\ln(|\lambda_{\mathrm{m}}-\lambda_{0}|/\sigma)$ for the SM
vs. delay time $\tau$ in the regime of strong chaos ($\sigma=12\,\mathrm{ns}^{-1}$).
Parameters as in Fig.~\ref{F1}.}
\end{figure}

Fig.~\ref{F2}(a) shows $\lambda_{\mathrm{m}}\,\tau$ as a function
of the delay time $\tau$ in the regime of weak chaos. We observe
that this product saturates at a constant value for large delay times.
Note that for our parameters a delay time of $200\,\mathrm{ns}$ is
much larger than the internal time scale ($1\,\mathrm{ns}$). For
the regime of strong chaos, Fig.~\ref{F2}(b) depicts $\ln(|\lambda_{\mathrm{m}}-\lambda_{0}|/\sigma)$
as a function of the delay time $\tau$. We observe that it decreases
linearly with $\tau$ in agreement with the analysis of Eq.~\eqref{E6}.

\begin{figure}
\includegraphics[height=0.34\columnwidth]{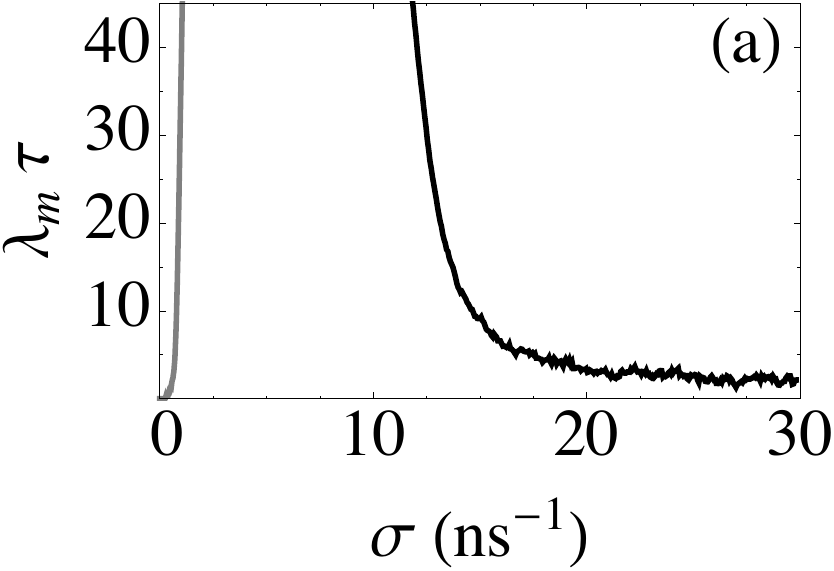}\hspace*{\fill}\includegraphics[height=0.34\columnwidth]{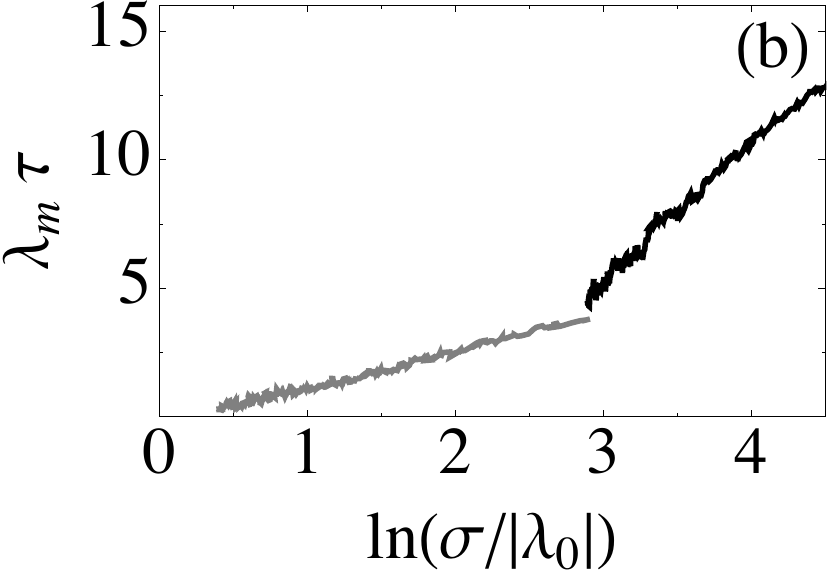}
\caption{\label{F3}(a)~$\lambda_{\mathrm{m}}\,\tau$ for the SM vs. coupling
strength $\sigma$ for $\tau=100\,\mathrm{ns}$. (b)~$\lambda_{\mathrm{m}}\,\tau$
for the SM vs. $\ln(\sigma/|\lambda_{0}|)$ in the regime of weak
chaos near the left (gray line) and right (black line) divergences.}
\end{figure}

At the transitions from weak to strong chaos, $\lambda_{\mathrm{m}}\,\tau$
diverges as shown in Fig.~\ref{F3}(a). In order to obtain the scaling
for this divergence, we first consider the simple case of a scalar
delay equation $\dot{s}=F(s)+b\, s(t-\tau)$ with constant coefficient
$b$. The corresponding characteristic equation of the fixed point
is
$\lambda_{\mathrm{m}}=\lambda_{0}+b\,\mathrm{e}^{-\lambda_{\mathrm{m}}\,\tau}$,
which can be solved using the Lambert-W function. For $\lambda_{0}\to0-$,
which corresponds to the transition from weak to strong chaos, it
predicts a divergence of $\lambda_{\mathrm{m}}\,\tau$ with $\ln(b/|\lambda_{0}|)$.
Assuming that the coupling strength $b$ in this simple model can
be identified with $\sigma$ in the chaotic LK equations, we observe
a similar scaling for the divergence of $\lambda_{\mathrm{m}}\,\tau$
at the two critical points of small and large values of $\sigma$.
Fig.~\ref{F3}(b) reveals that $\lambda_{\mathrm{m}}\,\tau$ indeed
depends linearly on $\ln(\sigma/|\lambda_{0}|)$. The slope is within
the same order of magnitude as for the simple case of steady states
but systematically larger. This deviation is related to the chaotic
time dependence of Eq.~\eqref{E3} and differs between the left (gray) and
right (black) divergences since the degree of chaotic fluctuations is different
for small and large $\sigma$. 

Up to now we have considered a single unit with time-delayed feedback
or, equivalently, the dynamics in the SM Eq.~\eqref{E2}. The stability
of chaos synchronization can be computed using the master stability
function \cite{Pecora1998}. It is defined
as the maximum LE $\lambda(r\,\mathrm{e}^{\mathrm{i}\,\psi})$ arising
from the variational Eq.~\eqref{E3} where $\sigma$ is replaced
by the complex parameter $\sigma\, r\,\mathrm{e}^{\mathrm{i}\,\psi}$
(the input trajectory $\mathbf{s}(t)$ is still governed by Eq.~\eqref{E2}).
For a given network the stability of the synchronized solution is
determined by the eigenvalues of $G$.

Due to the unity row sum the coupling matrix $G$ has one eigenvalue
$\tilde{\gamma}=1$ with eigenvector $(1,1,\dots,1)$, which corresponds
to perturbations in the SM. The other $N-1$ \emph{transversal} eigenvalues
$\gamma_{1},\dots,\gamma_{N-1}$ correspond to perturbations transversal
to the SM. Synchronization in the network is stable if $\lambda(\sigma\,\gamma_{k})<0$
for all transversal eigenvalues $\gamma_{k}$.

We now show that synchronization is stable for weak chaos if
\begin{equation}
|\gamma_{{\rm max}}|<\mathrm{e}^{-\lambda_{{\rm m}}\,\tau}\;,\label{E8}
\end{equation}
where $\gamma_{{\rm max}}$ is the transversal eigenvalue of $G$
with largest magnitude \cite{Englert2011}.

As follows from \cite{Flunkert2010}, in the large delay case, $\lambda(r\,\mathrm{e}^{\mathrm{i}\,\psi})$
does not depend on the phase $\psi$, and there exists a critical
value $r_{0}$ for the stability of the variational Eq.~\eqref{E3}
($\lambda(r_{0})=0$); i.\,e., for $r<r_{0}$, the perturbation $\boldsymbol{\xi}(t)$
vanishes asymptotically $\boldsymbol{\xi}(t)\to0$ and grows otherwise.
The maximum LE is zero for $r_{0}$. If the maximum LE $\lambda_{\mathrm{m}}$
for a given $\sigma$ is known, then the threshold $r_{0}$ can be
expressed as
\begin{equation}
r_{0}=\sigma\,\mathrm{e}^{-\lambda_{\mathrm{m}}\,\tau}\;.\label{E9}
\end{equation}
This can be shown by the following arguments. Let us make the coordinate
transformation $\boldsymbol{\xi}(t)=\mathbf{u}(t)\exp[t\ln(\sigma/r_{0})/\tau]$
in Eq.~\eqref{E3}. Then the variational equation in the transformed
coordinates reads
\begin{align}
\dot{\mathbf{u}}(t)= & \left(\mathrm{D}F[\mathbf{s}(t)]-\frac{1}{\tau}\ln\!\left(\frac{\sigma}{r_{0}}\right)I\right)\mathbf{u}(t)\nonumber \\
 & +r_{0}\,\mathrm{D}H[\mathbf{s}(t-\tau)]\,\mathbf{u}(t-\tau)\;.\label{E10}
\end{align}
The term $\ln(\sigma/r_{0})/\tau$ in the instantaneous part of the
vector field does not influence the maximum LE of Eq.~\eqref{E10}
in leading order $1/\tau$ for weak chaos. Indeed, by substituting
$\mathbf{u}\sim\mathrm{e}^{\mathrm{i}\,\omega\, t+\gamma/\tau}$ we
see that only the terms $\left|\dot{\mathbf{u}}\right|\sim\omega$,
$\left|\mathrm{D}F[s(t)]\,\mathbf{u}\right|\sim1$, as well as $\left|r_{0}\,\mathrm{D}H[s(t-\tau)]\,\mathbf{u}(t-\tau)\right|\sim r_{0}\,\mathrm{e}^{-\gamma}$
contribute to the leading order. Hence, the maximum LE of Eq.~\eqref{E10}
is zero, as well, and we obtain $\lambda_{\mathrm{m}}=\ln(\sigma/r_{0})/\tau$,
taking into account the relation between $\mathbf{u}$ and $\boldsymbol{\xi}$.
This leads to the estimate \eqref{E9} for the critical value $r_{0}$.
Then synchronization is stable if $\lambda(\sigma\,\gamma_{k})<0$,
and hence if $|\sigma\,\gamma_{k}|<r_{0}$ for all $k$. With Eq.~\eqref{E9}
this results in Eq.~\eqref{E8}.

The condition \eqref{E8} rules out synchronization for networks with
strong chaos since the right-hand side of Eq.~\eqref{E8} decreases
to zero in the limit of large delay times $\tau$. For weak chaos,
however, one can always find networks for which Eq.~\eqref{E8} is
true, i.\,e., chaos synchronization is stable. In addition, for weak
chaos, condition \eqref{E8} becomes independent of $\tau$, as $\lambda_{\mathrm{m}}\,\tau\to\mathrm{const}$,
in agreement with recent results \cite{Flunkert2010,Englert2011}.
The network can be synchronized in this case even for arbitrarily
large $\tau$ if Eq.~\eqref{E8} is fulfilled.

For a single laser with feedback we have found a scenario leading
from weak to strong chaos and back to weak chaos with increasing feedback
strength $\sigma$. But also for networks outside the regime of synchronization,
we can define an instantaneous LE for each unit by the maximum LE
of the equation
$\dot{\boldsymbol{\xi}}(t)=\mathrm{D}F\!\left[\mathbf{x}_{i}(t)\right]\boldsymbol{\xi}(t)$.
Simulating this together with Eq.~\eqref{E1} for a triangle of
bidirectionally coupled lasers, we found similar results as in Fig.~\ref{F1}.
The network changes from weak to strong chaos and back to weak chaos
with increasing coupling strength $\sigma$. The critical coupling
strengths, however, have different values.

Finally, we have performed an experiment with two coupled electronic
circuits \cite{Juengling2011} to measure the difference between strong
and weak chaos. 
\begin{figure}
\includegraphics[height=0.37\columnwidth]{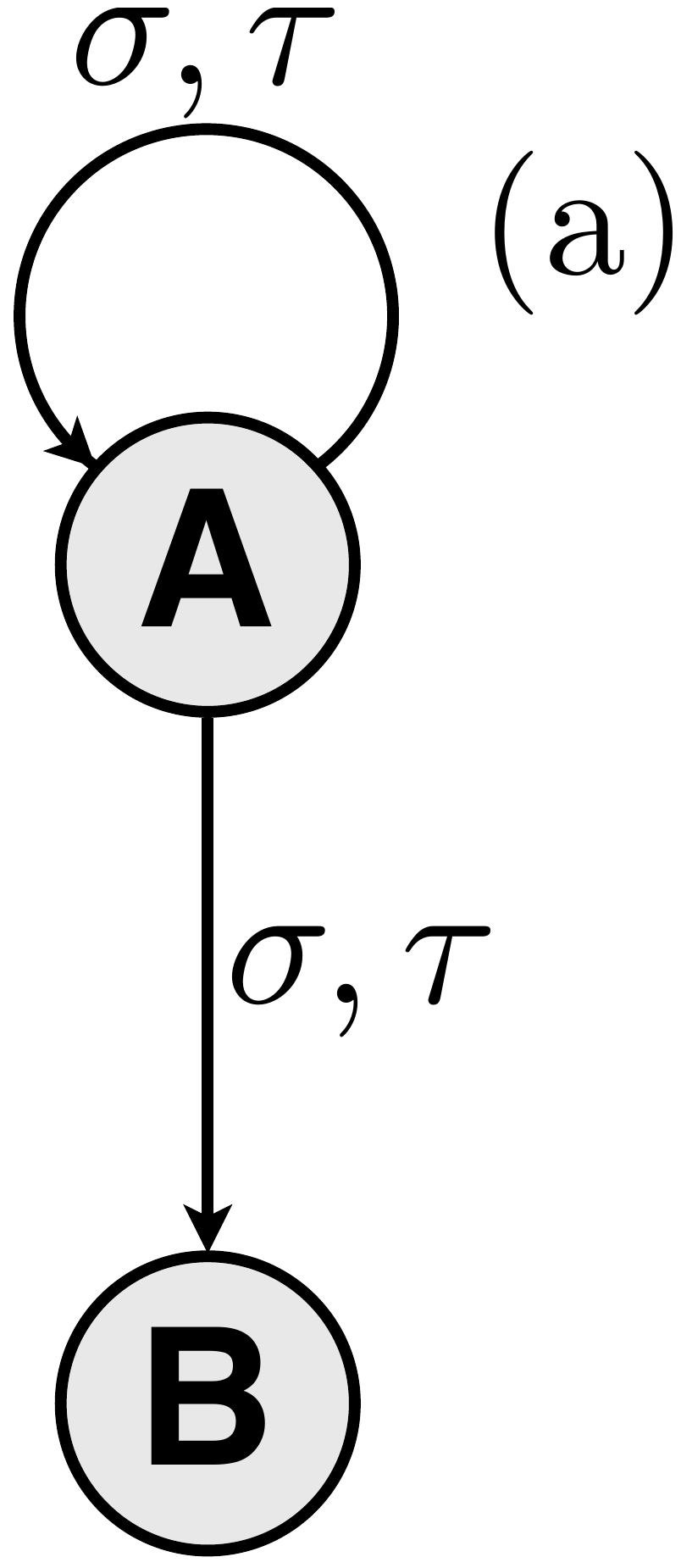}\hspace*{0.5cm}\includegraphics[height=0.37\columnwidth]{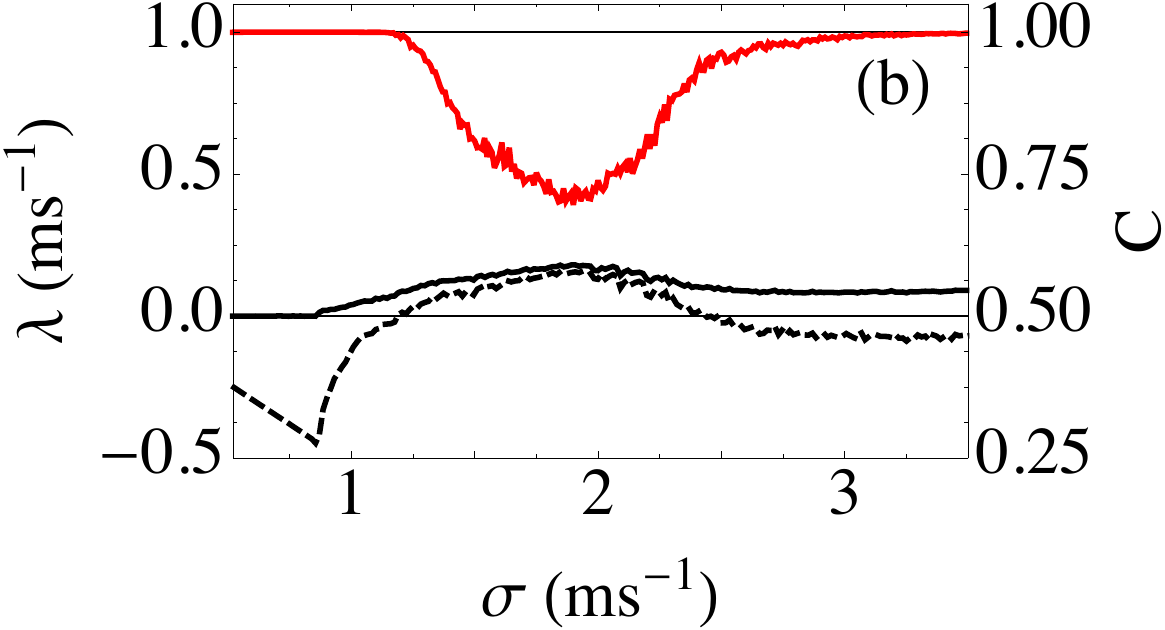}
\caption{\label{F4}(Color online) (a)~Experimental setup to measure the difference between
strong and weak chaos. (b)~Simulated $\lambda_{\mathrm{m}}$ (solid line) and 
$\lambda_{0}$ (dashed line) of the SM of the
two electronic circuits and experimentally measured cross-correlation $C$ (red (gray) line) 
between the maxima of the time series of the two electronic
circuits vs. coupling strength $\sigma$.}
\end{figure}
For general chaotic networks, one can add two identical units which
are driven by one unit of the network with identical strengths, similar
to the test for generalized synchronization \cite{Abarbanel2001}.
Chaos is weak if and only if the two units synchronize. For determining
the type of chaos on the SM, it is sufficient to add one unit which
is driven by a single unit with delayed feedback representing the
SM, as sketched in Fig.~\ref{F4}(a). The stability of synchronization
of the two units is given by Eq.~\eqref{E4}. Fig.~\ref{F4}(b)
shows the simulated LEs $\lambda_{\mathrm{m}}$ and $\lambda_{0}$
of the SM in comparison with the experimentally measured cross-correlation $C$
between the maxima of the time series of the two electronic circuits
as a function of the coupling strength $\sigma$. For small $\sigma$
we observe zero-lag synchronization of periodic dynamics. With increasing
$\sigma$ the dynamics becomes chaotic while complete synchronization
is maintained. With further increase of $\sigma$ the cross-correlation
first decreases and then increases again until synchronization is
reached once more, indicating transitions from weak to strong chaos
and back to weak chaos.

The notion of strong and weak chaos allows for a classification of
the synchronizability for coupled chaotic nodes and, most notably,
shows a significant difference of the chaotic behavior, characterized
by the maximum and the instantaneous LE. Our findings are promising for applications
in laser dynamics and beyond. In random number generators, where a
high entropy is crucial, the regime of strong chaos will potentially
lead to an increase in randomness. Similarly, in other applications
like opto-electronic oscillators, devices can be deliberately constructed
to operate in the regime of strong chaos.

\textbf{Acknowledgments} --- We thank the Deutsche Forschungsgemeinschaft
(SFB 910), the Leibniz-Rechenzentrum Garching and Hartmut
Benner from TU Darmstadt for their support of this work.


%

\end{document}